\pdfoutput=1	

\documentclass[10pt]{iopart}

\usepackage{hyperref}
\usepackage{graphicx}

\usepackage{rotating}

\begin{document}


\title{Local light-ray rotation}

\author{Alasdair C.\ Hamilton, Bhuvanesh Sundar, John Nelson, and Johannes Courtial}
\ead{j.courtial@physics.gla.ac.uk}
\address{Department of Physics and Astronomy, Faculty of Physical Sciences, University of Glasgow, Glasgow G12~8QQ, United~Kingdom}

\date{\today}

\begin{abstract}
We present a sheet structure that rotates the local ray direction through an arbitrary angle around the sheet normal.
The sheet structure consists of two parallel Dove-prism sheets, 
each of which flips one component of the local direction of transmitted light rays.
Together, the two sheets rotate transmitted light rays around the sheet normal.
We show that the direction under which a point light source is seen is given by a M\"{o}bius transform.
We illustrate some of the properties with movies calculated by ray-tracing software.
\end{abstract}




\section{Introduction}
Arrays of Dove prisms have been discussed as beam deflectors in contexts such as image stabilization and beam scanning \cite{Lian-Chang-1996}.
They have also been patented for use in a derotation assembly \cite{Watkins-2000}.
Ref.\ \cite{Lian-Chang-1996} also discusses combinations of two parallel Dove-prism arrays, one immediately behind the other, whereby the prism axes of the first array are perpendicular to those of the second.

We are interested in Dove-prism arrays consisting of very small Dove prisms, which we call Dove-prism sheets.
Without being aware of the combinations of parallel Dove-prism arrays discussed in Ref.\ \cite{Lian-Chang-1996}, we recently investigated the unusual imaging properties of two parallel Dove-prism sheets with crossed prism axes \cite{Courtial-Nelson-2008}.
Notably, in the ray-optical limit they behave like the interface between two optical media with equal and opposite refractive indices, and they form pseudoscopic images.

Here we generalize the concept to pairs of parallel Dove-prism sheets whose prism axes are rotated with respect to each other through an angle $\alpha/2$ around the sheet normal.
If the sheets are close together, their combined effect is a local ray rotation through an angle $\alpha$ around the common sheet normal.
Such ray-rotation sheets are examples of \underline{meta}ma\underline{t}erials f\underline{o}r ra\underline{y}s (METATOYs) -- called this because of a number of analogies with metamaterials in the original sense \cite{Smith-et-al-2004} -- which can, in the sense outlined in Ref.\ \cite{Hamilton-Courtial-2009}, perform ray-optics without wave-optical analog.
Local ray rotation through angles $\alpha$ other than $0^\circ$ and $180^\circ$ has no wave-optical analog; describing such purely ray-optical ray rotation in terms of the wave-optically motivated Fermat's principle leads to a formal description in terms of a generalized Snell's law that uses complex refractive indices \cite{Sundar-et-al-2009}.
We believe the unusual properties of ray-rotating sheets will find applications, perhaps in ranging, maybe in imaging \cite{Hamilton-Courtial-2008c}, most likely as toys.
We illustrate some of these properties with images and movies created with ray-tracing software.

This paper is organized as follows.
First we introduce the idea of local ray rotation by passage through parallel Dove-prism sheets;
then we establish and demonstrate the basic mathematics of ray rotation;
and finally we present an outlook on future work and conclude.


\begin{figure}
\begin{center}
\includegraphics{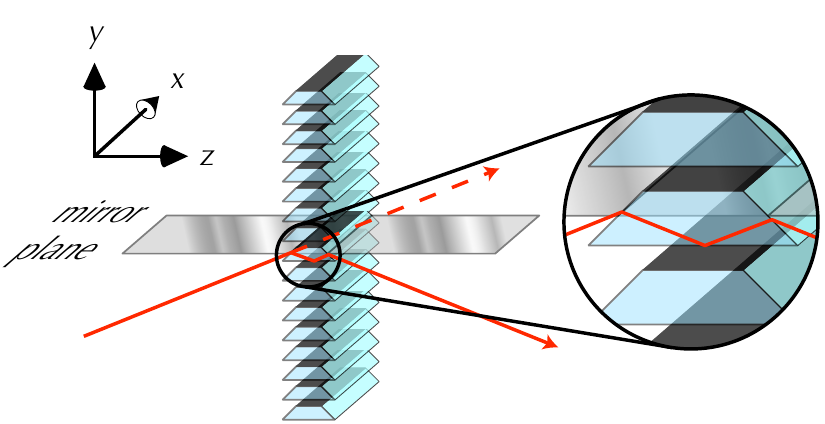}
\end{center}
\caption{\label{Dove-prism-sheet-figure}Ray-optics of a Dove-prism sheet.
With the coordinate system chosen as shown, the sheet inverts the $y$ component of any light ray (red) passing through it.
If the prism apertures are sufficiently small, the transverse shift of the light ray can be ignored.
The light ray's trajectory is then the original trajectory (dashed arrow), reflected with respect to a mirror plane normal to the sheet, parallel to the Dove prisms, and intersecting the sheet at the same point as the light ray.}
\end{figure}

\section{\label{dps-section}Dove-prism sheets and local ray rotation}
A Dove-prism sheet is an array of Dove prisms arranged as shown in Fig.\ \ref{Dove-prism-sheet-figure}.
In the coordinate system chosen as in figure \ref{Dove-prism-sheet-figure} the array inverts the $y$ direction of any transmitted light ray.
This is equivalent to mirroring the transverse ray direction with respect to a plane normal to the sheet and intersecting the sheet parallel to the Dove prisms.
The Dove-prism sheet also shifts the light rays in the transverse direction.
This shift is on the scale of the Dove-prism diameter, so as the Dove prisms are miniaturized, the shifts become increasingly negligible.
This miniaturization of optical elements can also be applied to confocal lenslet arrays, which then behave approximately like the interface between optical media with different refractive indices \cite{Courtial-2008a}.
Visually, the effect of viewing a straight line perpendicular to a Dove-prism sheet through the sheet distorts the line into a hyperbola \cite{Hamilton-Courtial-2008a}.

Clearly, the miniaturization of the Dove prisms must not be taken too far:
if the Dove-prism diameter get too close to, or below, the wavelength of the light it is designed for, diffraction will dominate over the light-ray-direction changes we are interested in here.
Acceptable compromises for visual purposes could be prism diameters of between $100 \mu$m and 1mm.
It is also worth mentioning that Dove prisms alter the polarization of transmitted light \cite{Padgett-Lesso-1999,Moreno-2004}, but this does not normally affect a Dove prism's visual performance.

\begin{figure}
\begin{center} \includegraphics{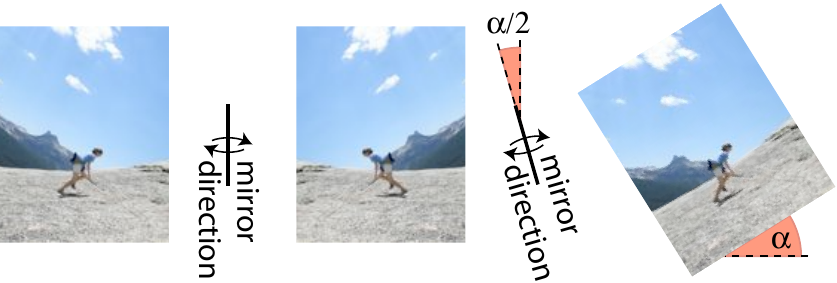} \end{center}
\caption{\label{rotation-figure}Mirroring and rotation.  In the example shown here, an image is, from left to right, first mirrored with respect to the vertical axis and then with respect to an axis at an angle $\alpha/2 = 16^\circ$ relative to the vertical.  The resulting image is rotated through $\alpha = 32^\circ$ relative to the original image.}
\end{figure}

Two identical Dove-prism sheets with the same orientation, but slightly displaced in the $z$ direction, mirror the local ray direction twice, with respect to the same plane.
These two reflections cancel each other, so such double-Dove-prism sheets transmit light rays without altering them (neglecting minor reflection, absorption and transverse-shift effects).
If the second sheet is rotated through an angle $\alpha/2$ around the $z$ axis, it mirrors the ray direction with respect to a plane that has been rotated with the sheet, and which is different from the first sheet's mirror plane.
Mirroring with respect to one plane, followed by mirroring with respect to another plane, leads to rotation through twice the angle between the planes (or lines, in the case of mirroring in two dimensions -- see Fig.\ \ref{rotation-figure}) around the line where the planes intersect.
In the case of two parallel Dove-prism sheets that are rotated with respect to each other around the sheet normal through an angle $\alpha/2$, this results in a rotation of the ray direction through an angle $\alpha$ around the sheet normal through the intersection point.
We call $\alpha$ the ray-rotation angle.

\begin{figure}
\begin{center} \includegraphics{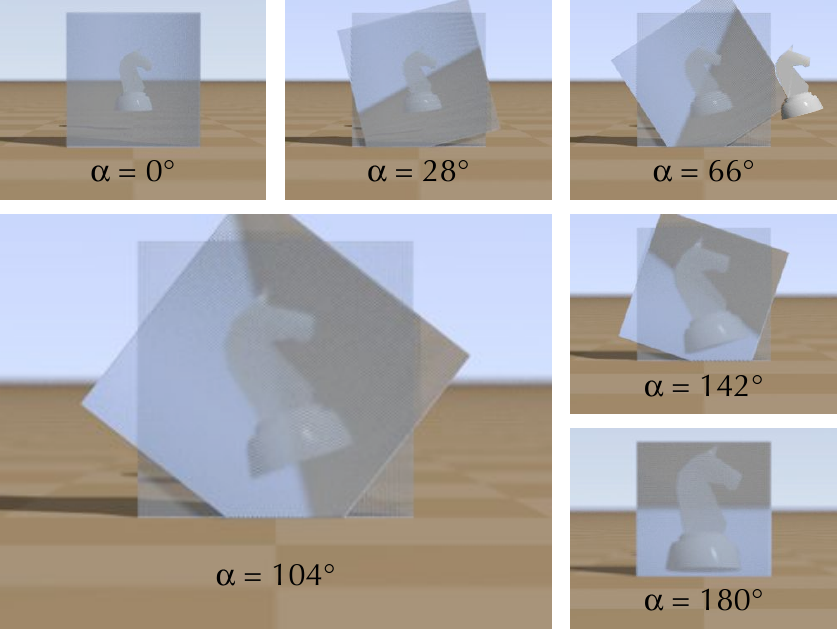} \end{center}
\caption{\label{vary-alpha-figure}Chess piece seen through ray-rotating Dove-prism sheets for various ray-rotation angles.
The Dove-prism sheets are rotated with respect to each other through an angle $\alpha/2$ ranging from $0^\circ$ to $90^\circ$, resulting in local ray rotation through an angle $\alpha$ between $0^\circ$ and $180^\circ$.
For $\alpha = 0$, the effect of the two Dove-prism sheets cancel each other out.  In this case, the Dove-prism sheet sandwich does not alter the ray direction.
$\alpha = 180^\circ$ is the case of crossed Dove-prism sheets discussed in Ref.\ \cite{Courtial-Nelson-2008}; the crossed Dove-prism-sheets image the chess piece pseudoscopically from a distance $z_1$ behind the sandwich to the same distance $z_1$ in front of the sheets.
For intermediate angles $\alpha$, the chess piece appears rotated and scaled.
The rotation angle and scale factor are described by Eqns (\ref{image-rotation-equation}) and (\ref{image-scaling-equation}); the additional chess piece in the $\alpha=66^\circ$ frame has been rotated and scaled accordingly.
The chess piece was positioned a distance of $z_1 = 2$ (in units of the side length of the floor tiles) behind the Dove-prism sheets, which were a distance $z_2 = 6$ in front of the camera.
The frames are from a movie (MPEG-4, 236 KB, available in the supporting online material) calculated by performing ray tracing through the detailed prism-sheet structure (each sheet contains 100 Dove prisms),
using the freely-available software POV-Ray~\cite{POV-Ray}.
}
\end{figure}

\begin{figure}
\begin{center} \includegraphics{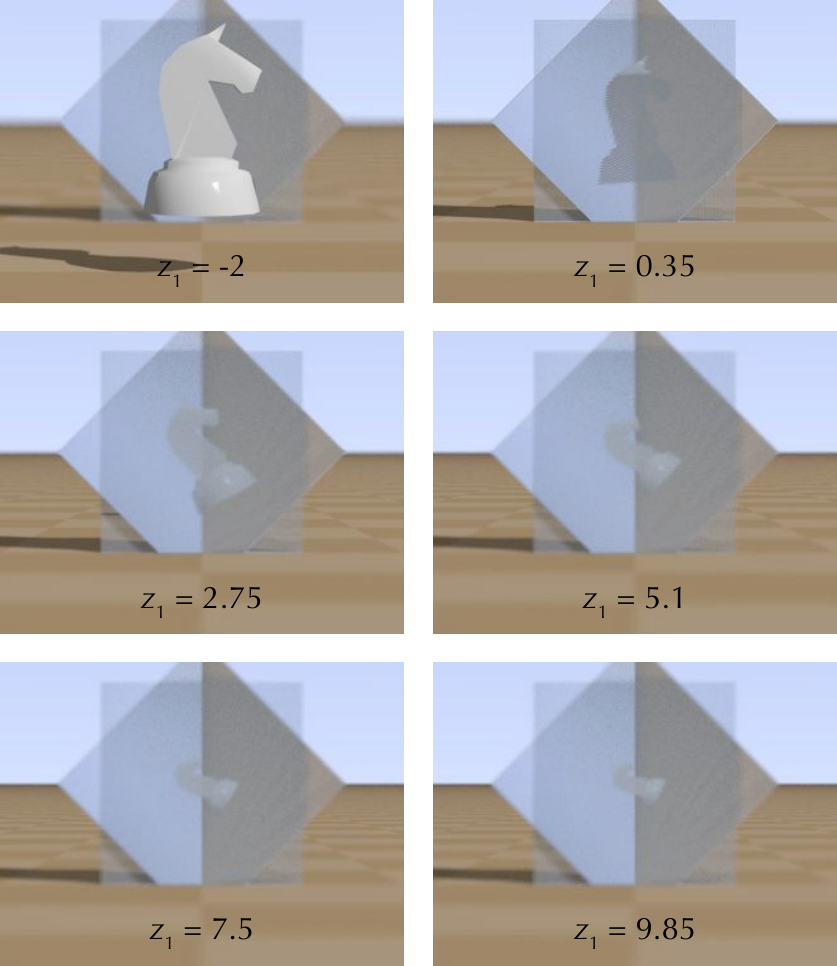} \end{center}
\caption{\label{90-degree-rotation-figure}Chess piece seen through Dove-prism sheets rotating the direction of transmitted light rays through $\alpha = 90^\circ$ around the sheet normal.
Different frames are calculated for various distances $z_1$ of the chess piece behind the sheets.
The distance between the sheet and the camera is $z_2 = 6$ (in units of the floor-tile side length).
The frames are taken from a movie (MPEG-4, 180 KB, available in the supporting online material) created with POV-Ray~\cite{POV-Ray}.}
\end{figure}

Fig.\ \ref{vary-alpha-figure} visualises a stationary object seen from a fixed position through two Dove-prism sheets for various ray-rotation angles $\alpha$.
For $\alpha = 0^\circ$ and $\alpha = 180^\circ$, the object appears upright, for all other angles it appears rotated slightly, whereby the rotation angle depends on $\alpha$.
Fig.\ \ref{90-degree-rotation-figure} keeps the ray-rotation angle fixed to $\alpha = 90^\circ$, but changes the distance between sheet and object (Fig.\ \ref{90-degree-rotation-figure}).
The object is again seen from a fixed position.
As the object moves away from the sheet, it appears to rotate.
The horizon -- an object at infinity -- appears rotated through $90^\circ$.
We investigate this apparent object rotation in more detail in the following section.

It is worth noting that the two special cases discussed above are the only rotation angles for which a ray-rotating sheet performs geometric imaging.
For $\alpha = 0^\circ$, light rays are neither rotated nor shifted; in terms of geometric imaging, the sheets produce an image at the same position as the object.
The case $\alpha = 180^\circ$ is that of the crossed Dove-prism sheets discussed in Ref.\ \cite{Courtial-Nelson-2008}.
The sheets then create a real image of the object at the same distance as the object, but on the opposite side of the sheet.
In the example shown in Fig.\ \ref{vary-alpha-figure}, the image is considerably closer to the camera and therefore appears bigger despite the fact that it is the same transverse size as the object (the transverse magnification is 1).
For all other ray-rotation angles $\alpha$, not all light rays originating from the same point light source on one side of the sheet intersect again in a (different) point after transmission through the sheet.
(The only plane that is always ``imaged'' is -- trivially -- the plane of the sheet, which is imaged into itself.)
The imaging properties of parallel ray-rotating sheets are investigated in Ref.\ \cite{Hamilton-Courtial-2008c}.

\section{\label{maths-section}Mathematics of ray rotation}
Here we derive mathematically the direction under which an observer would see a point on the other side of a ray-rotating sheet.

Consider a point light source $L$ a distance $z_1$ in front of a sheet that rotates the ray direction through an angle $\alpha$ around the local sheet normal.
The eye, $E$, is a distance $z_2$ in front of the sheet.
The point light source sends out light rays in all directions.
Many of these light rays pass through the ray-rotating sheet, where their transverse direction becomes rotated.  
If any of these light rays subsequently pass through the eye position, the eye sees the point light source in the direction these rays are coming from.

\begin{figure}
\begin{center} \includegraphics{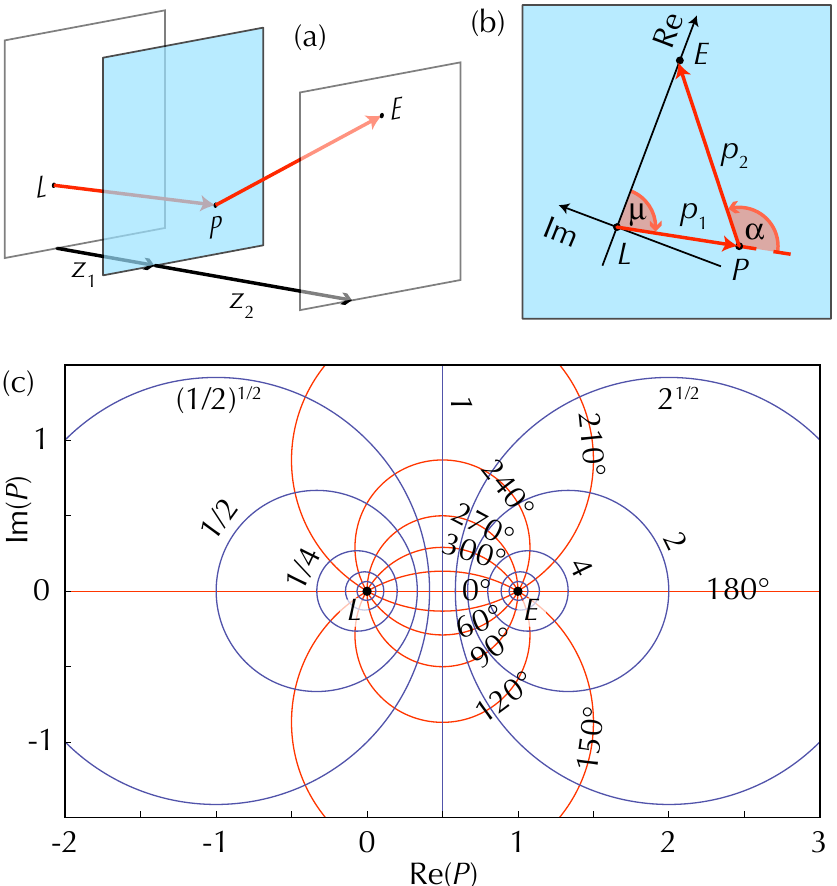} \end{center}
\caption{\label{projection-figure}Geometry of a light ray (red) originating at point light source $L$, hitting a ray-rotating sheet (blue) at point $P$, where it is rotated through an angle $\alpha$ around the sheet normal, and finally hitting the eye point, $E$.
(a)~Three-dimensional (3D) representation;
(b)~parallel projection into the sheet plane and definition of the $(\xi, \psi)$ coordinate system in the projection plane and geometry of the triangle with corners $L$, $P$, $E$;
(c)~family of curves in the $(\xi, \psi)$ coordinate system on which $P$ can lie, for various values of $r$ and $\alpha$.}
\end{figure}

We consider a light ray intersecting the sheet at point $P$ and passing through the eye point, $E$ (Fig.\ \ref{projection-figure}(a)).
We study the light ray's orthographic projection into the plane of the sheet (Fig.\ \ref{projection-figure}(b)).

We notice that the angle between the \emph{projections} of the ray in front of and behind the sheet is precisely the angle $\alpha$ through which the sheet rotates the transverse ray direction.
What about the lengths of the projections?
Initially the light ray is travelling at an angle $\theta$ with respect to the sheet normal.
From fundamental trigonometry we see that the length $p_1$ of the projection of the ray between $L$ and $P$ is related to the distance between $L$ and the sheet, $z_1$, through the equation
\begin{equation}
\frac{p_1}{z_1} = \tan \theta.
\label{tan-1-equation}
\end{equation}
But because the sheet rotates the ray direction around the sheet normal, the angle of the transmitted light ray relative to that sheet normal is still $\theta$.
This means that the length $p_2$ of the projection of the ray between $P$ and $E$ is related to the distance between the sheet and $E$, $z_2$, through
\begin{equation}
\frac{p_2}{z_2} = \tan \theta.
\label{tan-2-equation}
\end{equation}
Eliminating $\tan \theta$ from equations (\ref{tan-1-equation}) and (\ref{tan-2-equation}) 
and re-arranging gives
\begin{equation}
\frac{p_2}{p_1} = \frac{z_2}{z_1} = r. 
\label{r-equation}
\end{equation}

The distance ratio $r$ and the ray-rotation angle $\alpha$ then completely determine the position of $P$ relative to the projections of $L$ and $E$.
In the following $L$, $P$ and $E$ refer to the projections into the sheet plane of the three-dimensional positions $L$, $P$ and $E$.
We place a complex plane into the projection plane.
$L$, $P$, and $E$ are then represented by complex numbers.
We position, scale and orientate the complex plane such that $L = 0$ and $E = 1$ (Fig.\ \ref{projection-figure}(b)).

Equation (\ref{r-equation}) can then be written as
\begin{equation}
\frac{|E-P|}{|P|} = r,
\end{equation}
and because the angle between the complex-plane vectors $(E-P)$ and $P$ is $\alpha$,
\begin{equation}
E-P = r \exp(i \alpha) P.
\end{equation}
Then
\begin{equation}
1 = E = P + (E-P) = P (1+r \exp(\rmi \alpha)),
\end{equation}
and so
\begin{equation}
P = \frac{1}{1+r \exp(\rmi \alpha)}.
\label{P-equation}
\end{equation}
Figure \ref{projection-figure}(c) shows the lines of constant $\alpha$ and $r$.

We note that Eqn 
(\ref{P-equation}) is the M\"{o}bius transformation \cite{Needham-2000-Moebius-transformation} 
(otherwise known as a linear fractional transformation \cite{Weisstein-2002}) $(a z + b) / (c z + d)$ with $z = r \exp ( \rmi \alpha )$, $a = 0$ and $b = c = d$.
The lines of constant $\alpha$ and the lines of constant $r$ also form a bipolar coordinate system~\cite{Wikipedia-bipolar-coordinates-2008}.

\begin{figure}
\begin{center} \includegraphics{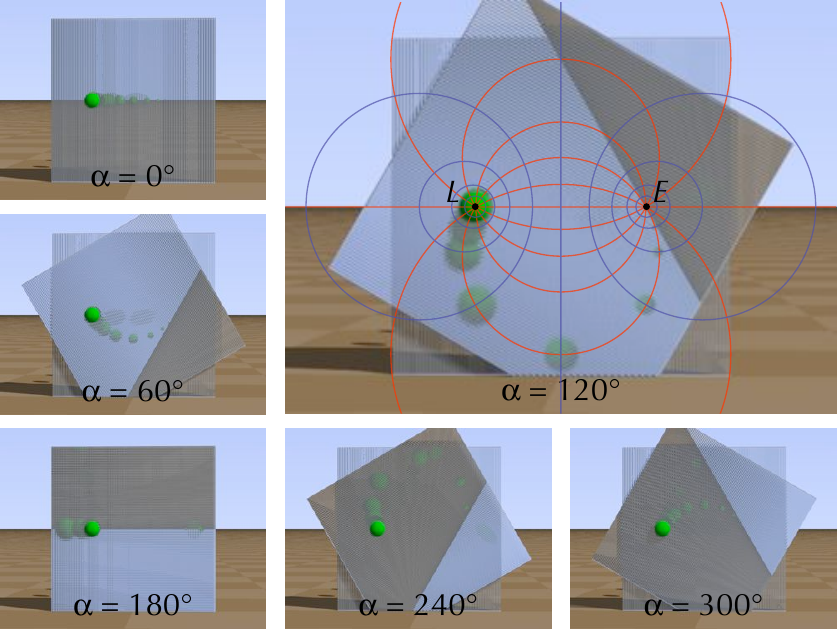} \end{center}
\caption{\label{Moebius-ray-tracing-figure}Dependence of the apparent position of objects seen through a ray-rotating sheet ($2 \times 100$ Dove prisms) on the ray-rotation angle, $\alpha$, and the ratio between the distances between the object and the sheet and between the observer and the sheet, $r$.
The frames show the apparent positions of several green spheres, all positioned along a sheet normal through the point $L$ at various distances from the sheet.
Different frames are calculated for different ray-rotation angles, $\alpha$.
The green spheres play the role of light sources $L$ located different distances $z_1$ behind the sheet.
In terms of the distance between the camera and the sheet, $z_2$, these distances are  $z_1 = 2^n z_2$, $n = -5, ..., 5$.
The camera is positioned on a sheet normal through $E$, at a distance $z_2 = 6$ (in units of floor-tile side lengths).
The frames are taken from a movie (MPEG-4, 326 KB, available in the supporting online material) created using the ray-tracing software POV-ray~\cite{POV-Ray}.}
\end{figure}

Fig.\ \ref{projection-figure}(c) shows the position of $P$ in the $(\xi, \psi)$ coordinate system for various values of $r$ and $\alpha$.
The dependence of $P$ on $r$ and $\alpha$ is that expected from bipolar coordinates:
the lines of constant $r$ and the lines of constant $\alpha$ are circles intersecting each other at right angles \cite{Wikipedia-bipolar-coordinates-2008} (Appolonian circles \cite{Wikipedia-Apollonian-circles-2008}).
If $r$ is varied and $\alpha$ is kept constant, $P$ moves on circle segments connecting $L$ and $E$, whereby $L$ corresponds to $r = 0$ and $E$ corresponds to the limit $r \rightarrow \infty$;
if $\alpha$ is varied and $r$ is kept constant, $P$ moves on circles surrounding (but not centered on) either $L$ or $E$, depending on the value of $r$.
This dependence of the apparent position $P$ on $r$ and $\alpha$ is demonstrated in Fig.\ \ref{Moebius-ray-tracing-figure}.
It can also be confirmed by careful tracing of the apparent position of specific points on the object in the different frames in figures \ref{vary-alpha-figure} and \ref{90-degree-rotation-figure}.





\section{\label{image-rotation-section}Apparent rotation of extended objects}
Light from a point light source at position $L$ arrives from apparent position $P$, which is rotated relative to the actual position, $L$,  by the angle $\nu$ around the projection of the eye position, $E$ (Fig.\ \ref{projection-figure}(b)). 
As can be seen from Fig.\ \ref{projection-figure}(b), this angle is the argument of the complex number $P$, so
\begin{equation}
\nu = \arg(P) = \arg \left( \frac{1}{1+r \exp(\rmi \alpha)} \right).
\label{image-rotation-equation}
\end{equation}
$\nu$ is a function of $r$ and $\alpha$ only, which means that, when seen from a constant eye position (constant $z_2$), it is the same for all light sources in the same transverse plane (same $z_1$).
This in turn implies that any extended object in a transverse plane appears rotated as a whole.
It is also magnified; the magnification (relative to the same object in the sheet plane) is
\begin{equation}
M = |P| = \left| \frac{1}{1+r \exp(\rmi \alpha)} \right|.
\label{image-scaling-equation}
\end{equation}
The results from these equations agree with our POV-Ray simulations; an example of a chess piece that has been rotated and scaled according to Eqns (\ref{image-rotation-equation}) and (\ref{image-scaling-equation}) is shown in one of the frames in Fig.\ \ref{vary-alpha-figure}.



\section{\label{conclusions-section}Conclusions} 
In this paper we have described a sheet structure that rotates the direction of transmitted light rays through a fixed angle around the sheet normal.
This leads to very unusual optical effects such as the apparent rotation of extended objects when 
seen through a ray-rotating sheet, which we have also described mathematically. 

Currently, we are working on a generalization of local light-ray rotation to rotation axes with arbitrary directions \cite{Hamilton-et-al-2009b}.
We are also in the process of realizing local light-ray rotation experimentally.
As a first step, we recently built sheets equivalent to Dove-prism sheets from lenticular arrays \cite{Blair-et-al-2009}.
The optical quality of our sheets is not yet sufficient to combine them into convincing ray-rotation sheets, but we are working on improvements to these sheets.


\ack
Thanks to Tom\'a\v{s} Tyc for stimulating discussions.
ACH and JN are supported by the UK's Engineering and Physical Sciences Research Council (EPSRC).
JC is a Royal Society University Research Fellow.

\section*{References}

\bibliographystyle{osajnl}
\bibliography{/Users/johannes/Documents/work/library/Johannes}

\begin{thebibliography}{10}
\newcommand{\enquote}[1]{``#1''}

\bibitem{Lian-Chang-1996}
T.~Lian and M.-W. Chang, \enquote{New types of reflecting prisms and reflecting
  prism assembly,} Optical Engineering \textbf{35}, 3427--3431 (1996).

\bibitem{Watkins-2000}
R.~A. Watkins, \enquote{Multiple dove prism assembly,} U. S. Patent {6,097,554}
  (2000).

\bibitem{Courtial-Nelson-2008}
J.~Courtial and J.~Nelson, \enquote{Ray-optical negative refraction and
  pseudoscopic imaging with {D}ove-prism arrays,} New J. Phys. \textbf{10},
  {023028} (2008).

\bibitem{Smith-et-al-2004}
D.~R. Smith, J.~B. Pendry, and M.~C.~K. Wiltshire, \enquote{{Metamaterials and
  Negative Refractive Index},} Science \textbf{305}, 788--792 (2004).

\bibitem{Hamilton-Courtial-2009}
A.~C. Hamilton and J.~Courtial, \enquote{Metamaterials for light rays: ray
  optics without wave-optical analog in the ray-optics limit,} New J. Phys.
  \textbf{11}, 013042 (2009).

\bibitem{Sundar-et-al-2009}
B.~Sundar, A.~Hamilton, and J.~Courtial, \enquote{Fermat's principle with
  complex refractive indices and local light-ray rotation,} Opt. Lett.
  \textbf{34}, 374--376 (2009).

\bibitem{Hamilton-Courtial-2008c}
A.~C. Hamilton and J.~Courtial, \enquote{Imaging with parallel ray-rotation
  sheets,} Opt. Express \textbf{16}, 20826--20833 (2008).

\bibitem{Courtial-2008a}
J.~Courtial, \enquote{Ray-optical refraction with confocal lenslet arrays,} New
  J. Phys. \textbf{10}, 083033 (2008).

\bibitem{Hamilton-Courtial-2008a}
A.~C. Hamilton and J.~Courtial, \enquote{Optical properties of a {D}ove-prism
  sheet,} J. Opt. A: Pure Appl. Opt. \textbf{10}, 125302 (2008).

\bibitem{Padgett-Lesso-1999}
M.~J. Padgett and J.~P. Lesso, \enquote{{D}ove prisms and polarised light,} J.
  Mod. Opt. \textbf{46}, 175--179 (1999).

\bibitem{Moreno-2004}
I.~Moreno, \enquote{Jones matrix for image-rotation prisms,} Appl. Opt.
  \textbf{43}, 3373--3381 (2004).

\bibitem{POV-Ray}
\enquote{{POV-Ray -- The Persistence of Vision Raytracer},}
  http://www.povray.org/.

\bibitem{Needham-2000-Moebius-transformation}
T.~Needham, \emph{Visual Complex Analysis} (Clarendon Press, 2000), chap.~3.

\bibitem{Weisstein-2002}
E.~W. Weisstein, \enquote{Linear fractional transformation,} MathWorld (2002).

\bibitem{Wikipedia-bipolar-coordinates-2008}
Wikipedia, \enquote{{Bipolar coordinates},}  (2008). Accessed 3/7/2008.

\bibitem{Wikipedia-Apollonian-circles-2008}
Wikipedia, \enquote{{Apollonian circles},}  (2008). Accessed 3/7/2008.

\bibitem{Hamilton-et-al-2009b}
A.~C. Hamilton, B.~Sundar, and J.~Courtial, \enquote{Local light-ray rotation
  around arbitrary axes,} in preparation (2009).

\bibitem{Blair-et-al-2009}
M.~Blair, L.~Clark, E.~A. Houston, G.~Smith, J.~Leach, A.~C. Hamilton, and
  J.~Courtial, \enquote{Experimental demonstration of a
  light-ray-direction-flipping {METATOY} based on confocal lenticular arrays,}
  arXiv:0902.3192 [physics.optics] (2009).

\end{thebibliography}

\end{document}